\documentclass[twocolumn,english]{IEEEtran}

\usepackage{amsfonts}
\usepackage{amsmath}
\usepackage{amssymb}
\usepackage{graphicx}
\usepackage{float}
\usepackage[mathscr]{eucal}
\usepackage{amsfonts}
\usepackage{float}
\usepackage[colorlinks,linkcolor=blue]{hyperref}
\usepackage[latin1]{inputenc}
\usepackage{amsmath}
\usepackage{graphicx}
\usepackage{amssymb}
\usepackage{amsthm}
\usepackage{verbatim}
\usepackage{epsfig}
\usepackage[labelfont=bf]{caption}
\usepackage{enumerate}
\usepackage{subfig}
\usepackage{epstopdf}

\newtheorem{lemma}{Lemma}
\newtheorem{proposition}{Proposition}
\newtheorem{corollary}{Corollary}
\newtheorem{fact}{Fact}

\newtheorem{remark}{Remark}

\newtheorem{assumption}{Assumption}

\def\begcen{\begin{center}}
\def\endcen{\end{center}}

\newcommand{\col}{ \mbox{col} }
\newcommand{\rank}{ \mbox{rank } }

\def\caly{{\cal Y}}
\def\calg{{\cal G}}

\def\calj{{\cal J}}

\def\calj{{\cal J}}

\def\liminf{\lim_{t \to \infty}}

\def\L2{{\cal L}_2}
\def\L2e{{\cal L}_{2e}}

\def\rea{\mathbb{R}}

\def\adj{\mbox{adj}}

\def\x{{x}}

\def\begmat#1{\begin{bmatrix}#1\end{bmatrix}}
\def\begali#1{\begin{align}{#1}\end{align}}
\def\begalis#1{\begin{align*}{#1}\end{align*}}

\def\begequarr{\begin{eqnarray}}
\def\endequarr{\end{eqnarray}}
\def\begequarrs{\begin{eqnarray*}}
\def\endequarrs{\end{eqnarray*}}
\def\begarr{\begin{array}}
\def\endarr{\end{array}}
\def\begequ{\begin{equation}}
\def\endequ{\end{equation}}
\def\lab{\label}
\def\begdes{\begin{description}}
\def\enddes{\end{description}}
\def\begenu{\begin{enumerate}}
\def\begite{\begin{itemize}}
\def\endite{\end{itemize}}
\def\endenu{\end{enumerate}}

\def\lef[{\left[\begin{array}}
\def\rig]{\end{array}\right]}

\def\begcen{\begin{center}}
\def\endcen{\end{center}}
\def\begrem{\begin{remark}\rm}
\def\endrem{\end{remark}}
\def\begassum{\begin{assumption}}
\def\endassum{\end{assumption}}
\def\begassums{\begin{assumption*}}
\def\endassums{\end{assumption*}}
\def\begassu{\begin{ass}}
\def\endassu{\end{ass}}
\def\beglem{\begin{lemma}}
\def\endlem{\end{lemma}}
\def\begcor{\begin{corollary}}
\def\endcor{\end{corollary}}
\def\begfac{\begin{fact}}
\def\endfac{\end{fact}}

\def\liminf{\lim_{t \to \infty}}

\def\L2e{{\cal L}_{2e}}

\def\rea{\mathbb{R}}

\def\adj{\mbox{adj}}
\def\col{\mbox{col}}

\def\et{\varepsilon_t}

\def\rank{\mbox{rank}\;}

\def\begsubequ{\begin{subequations}}
	\def\endsubequ{\end{subequations}}
\def\begpro{\begin{proposition}}
	\def\endpro{\end{proposition}}
\def\beglem{\begin{lemma}}
	\def\endlem{\end{lemma}}
\def\begass{\begin{assumption}}
	\def\endass{\end{assumption}}
\def\begcor{\begin{corollary}}
	\def\endcor{\end{corollary}}
\def\begproo{\begin{proof}}
	\def\endproo{\end{proof}}

\usepackage{xcolor}

\title{State Observer for the Fourth-order Model of a Salient Pole Synchronous Generator with Stator Losses: Known and Partially Unknown Input Cases}
	\author{Alexey Bobtsov, Romeo Ortega, Nicolai Lorenz-Meyer and Johannes Schiffer
	\thanks{A. Bobtsov is with Hangzhou Dianzi University (HDU), Hangzhou, China and ITMO University, Saint-Petersburg, Russian Federation (e-mail: bobtsov@mail.ru).}
	\thanks{R. Ortega is with Departament of Electrical and Electronic Engineering, Instituto Tecnológico Autónomo de México, Ciudad de México, Mexico and ITMO University, Saint Petersburg, Russian Federation (e-mail: romeo.ortega@itam.mx).}
		\thanks{N. Lorenz-Meyer is with Brandenburg University of Technology Cottbus-Senftenberg, 03046 Cottbus, Germany (e-mail: lorenz-meyer@b-tu.de).}
	\thanks{J. Schiffer is with Brandenburg University of Technology Cottbus-Senftenberg, 03046 Cottbus, Germany and Fraunhofer IEG, Fraunhofer Research Institution for Energy Infrastructures and Geothermal Systems, 03046 Cottbus, Germany (e-mail: schiffer@b-tu.de).}
}
\begin{document}

\maketitle
\begin{abstract}
In this paper we study the question of how to reconstruct the {\em state} of a power system using Phasor Measurement Units (PMUs). In our previous research we proved that this question has an affirmative answer imposing some rather strict structural assumptions: namely, neglecting the generator rotors saliency and assuming that the stator resistance of the synchronous generator is zero. It was shown in simulations that the performance of the proposed observer was sensitive to these assumptions, observing a transient quality degradation for realistic simulations not imposing these assumptions. Moreover, it was assumed in our previous work that the mechanical power and the field voltage are available for measurement, a scenario that it is not always realistic. In this paper we accomplish two ambitious objectives. First, we propose a new  observer that does not impose the simplifying assumptions on the generator model. Secondly, we consider the more realistic scenario where only mechanical power is available for measurement. That is, we solve a problem of state reconstruction of a nonlinear system with partially known input measurements---that is well-known to be a very challenging task. The design of the first observer relies on two recent developments proposed by the authors, a parameter estimation based approach to the problem of state estimation and the use of the Dynamic Regressor Extension and Mixing (DREM) technique to estimate these parameters. The use of DREM allows us to overcome the problem of lack of persistent excitation that stymies the application of standard parameter estimation designs. On the other hand, the observer for the partial input measurement scenario relies on the clever exploitation of the systems model. Simulation results illustrates the good performance of the proposed observers. 
\end{abstract}

.
\IEEEpeerreviewmaketitle

\section{Introduction}
\lab{sec1}
%
Driven by climate change mitigation measures paired with technological advancements, power systems world-wide are currently undergoing major transformations \cite{milano18,gu2022}. In particular, this comprises the large-scale integration of power-electronics interfaced renewable energy sources and complex loads, which are deployed in different hierarchical levels of power systems \cite{milano18,gu2022}. As a consequence, modern power systems exhibit higher and reverse power flows, leading to faster and more volatile system dynamics, which result more frequently in an operation closer to the stability limit \cite{milano18,gu2022}. 

The developments outlined above imply that an accurate monitoring of the system states becomes increasingly important for power system operators \cite{zhao2019,zhao2020}. Power system state estimation has always played an important role in power system monitoring, control
and protection \cite{zhao2020,liu2021}.  However, the predominant approaches used in today's power systems operation are steady state estimation (SSE) tools \cite{zhao2019}. As the denomination indicates, all SSE applications are developed under the assumption that the power system is operated at a {\em steady state}, giving rise to the conclusion that its behavior can be fully described by algebraic relations \cite{zhao2019}. Yet, it is obvious that this traditional steady-state assumption becomes more and more questionable with the transformation to modern climate-neutral power systems with increasingly volatile dynamics \cite{zhao2019}. This fact has been recognized by the power systems community and led to a swiftly emerging interest in dynamic state estimation (DSE) tools \cite{zhao2019,zhao2020}. This interest has been further substantiated by the increasing availability of phasor measurement units (PMUs) and advanced communication infrastructure \cite{zhao2019}. The following three primary potential application areas for DSE in power systems operation have been identified in \cite{zhao2020}: 1) model calibration and validation, 2) monitoring, and 3) operation. 

As the dynamics of synchronous generators (SGs) play a critical role in power systems \cite{machowski_power_2008}, a significant amount of the DSE literature considers the problem of estimating the states of SGs using locally available PMU-measurements \cite{zhao2019}. 
Most of the approaches are based on the Kalman filter (KF) and its extensions, such as extended KFs \cite{ghahremani_local_2016}, unscented KFs \cite{wang_alternative_2012}, and particle filters \cite{emami_particle_2015}.
In \cite{paul_dynamic_2018}, the states of a fourth-order SG model are estimated along with the exciter voltage and the mechanical torque utilizing an extended KF and placing the classical assumption of PMU-measurements available at the terminal bus of the SG. An adaptive unscented KF for a sixth-order SG model is proposed in the recent work \cite{hou_adaptive_2024}, which takes cyberattack-related measurement disturbances into account.
We refer the reader to \cite{singh_dynamic_2018,zhao2019,yang_review_2021} and the references therein for a thorough review of the literature on KF-based DSE techniques.

A key limitation of the KF-based works is that the convergence of the estimation algorithms is only demonstrated by simulation, without providing any formal stability guarantees. However, recently novel DSE approaches building on observer design techniques from a control systems background and providing analytic convergence guarantees for the estimation error have been proposed, see,  {\em e.g.}, \cite{qi_comparing_2018,nugroho_robust_2020,anagnostou_observer-based_2018}. An observer for excitation control for transient stabilization of a SG is proposed in \cite{rojas_observer-based_2020} with the assumption of the rotor angle being measurable, which may be difficult to fulfill in a practical setting. In \cite{taha_risk_2018}, a sliding mode observer is used as part of a risk mitigation technique to eliminate threats from cyber-attacks and unknown inputs utilizing a linearised power system model.

In \cite{LORetal}, the authors provided a globally convergent solution to the state observation problem for the case when the SG are modeled by the classical {\em third order} flux-decay model, which has been validated in \cite{lorenz2023} with real-world PMU measurement data of a power plant in the 500-1000 MW class within the extra-high voltage grid (410 kV) of Germany. 
This result was later extended in \cite{BOBetaltac22} to the case of a fourth order model that, as it is widely recognized \cite{MACetal}, gives a better description of the SG dynamics. In both works two critical assumptions regarding the generator model are made:
\begenu[{\bf (i)}]
\item the transient saliency is {\em neglected}, leading to the assumption that the direct- and quadrature axis transient reactances {\em are equal};
\item it is assumed that the SG {\em stator resistance is zero}.
\endenu 
As will be shown in the sequel, both assumptions considerably simplify the model of the SG, leading to a simpler observer design. On the other hand, via simulations it was shown in  \cite{LORetal} that the proposed observer is quite sensitive to the violation of these assumptions.  

In this paper we provide {\em two solutions} to the state observation problem for multimachine power systems described by the fourth order model with {\em salient} and {\em lossy} generators---that is {\em removing} the two aforementioned assumptions made in \cite{LORetal,BOBetaltac22}. The first solution assumes that {\em both input signals} are available for measurement. In the second solution we adopt the more practically reasonable scenario that {\em only the mechanical power} $P_m$ is measurable. For both observer designs---full and partial input measurements---we rely on a first {\em key step} of estimation of the shaft speed and derivation of a simple parameterization of the rotor angle, presented in Lemma \ref{lem1}. It should be mentioned that this important step was obviated in \cite{LORetal,BOBetaltac22} because of the simplifying assumptions on the SG model. 

The full input measurement observer relies on the direct utilisation of a suitable parameterization of the remaining state coordinates and the use of the measurement of the terminal current magnitude to derive a regression equation that can be used in the parameter estimator. Unfortunately, this is a nonlinearly parameterized regression equation (NLPRE) a difficulty that is bypassed using the new least-squares plus dynamic regression extension and mixing (LS+DREM) estimator of \cite{ORTROMARA}---that can handle NLPRE. An additional advantage of LS+DREM estimators is that the {\em excitation requirements} to ensure (exponential) parameter and estimated state convergence are minimal. Namely, it is required that the regressor satisfies an interval excitation (IE) condition \cite{KRERIE}. 

The partial input measurement observer, on the other hand, is derived ingeniously manipulating the dynamic equation of the quadrature axis internal model to estimate, via some suitable filtering and algebraic operations, the rotor angle---from which the remaining states can be directly derived. It is interesting to note that, in contrast with the previous observer, this observer design does not require the implementation of a parameter estimator. Therefore, no assumption on signal excitation is required.  

In the theoretical part of the paper we restrict ourselves to the study of a single generator. As shown in \cite{LORetal}, thanks to the incorporation of the PMUs, for the purposes of observer design it is possible to treat multimachine systems as a set of decentralized single machines, hence our result can be extended in a straightforward way to the multimachine case. In the interest of brevity we omit the details of this generalization, and refer the interest reader to \cite[Section II]{LORetal} for the details.  However, the simulation results include the multimachine case, and show the improved performance of the proposed observer with respect to a locally stable gradient-descent based observer. 

The rest of the paper has the following structure. The derivation of the SG model and explanation of the PMU signals are given in Section \ref{sec2}. In Section \ref{sec3} we present the first state observer, assuming all inputs are measurable. The observer with partial input measurements is given in  Section \ref{sec4}. Some simulation results are presented  in Section \ref{sec5}. Finally, in Section \ref{sec6} we include some concluding remarks.
%
\section{Synchronous Generator Model and PMU Measurements}
\lab{sec2}
In this section, the considered SG model is introduced. Thereby and as outlined in Section~\ref{sec1}, we focus on a single machine model. 

\subsection{Two-Axis Fourth-Order Model of a Synchronous Generator}
\lab{subsec21}
%
We consider the well-known two-axis fourth-order model of a SG with state vector and input vectors, see \cite[Equations (6.110-6.113)]{sauer2017}, \cite{MACetal},
$$
x=\begin{bmatrix}\delta & \omega & E_q' & E_d'\end{bmatrix},\quad u=\begin{bmatrix}P_m & E_f\end{bmatrix},
$$
where $\delta(t)\in[0,2\pi)$ denotes the rotor angle, $\omega(t)\in\rea$ the shaft speed relative to the synchronous speed $\omega_0\in\rea$, $E_d'(t)\in\rea$ and $E_q'(t)\in\rea$ the direct and quadrature axis internal voltages, respectively, and $P_m(t)\in\rea$ the mechanical power and $E_f(t)\in\rea$ the field voltage. 

By introducing the positive real constants $H,$ $D,$ $T_{d0}'$, $T_{q0}'$, $X_d,$ $X_d'$, $X_q$ and $X_q'$, the machine dynamics are given by
\begin{equation}
	\begin{split}
		\dot x_1&=x_2,\\
		\dot x_2&=\frac{\omega_0}{2H}\left(u_1-P_e-D x_2 \right),\\
		\dot x_3&=\frac{1}{T_{d0}'}\left(u_2-x_3-(X_d-X_d')I_d \right),\\
		\dot x_4&=\frac{1}{T_{q0}'}\left(-x_4+(X_q-X_q')I_q \right),
	\end{split}
\label{SG}
\end{equation}
with the electric air-gap power $P_e(t)\in\rea$ given by
\begin{equation}
P_e=E_q' I_q+E_d'I_d+(X_q'-X_d')I_dI_q,
\label{Pe}
\end{equation} 
where $I_q(t)\in\rea$ and $I_d(t)\in\rea$ denote the quadrature and direct axis currents, respectively. The last term in \eqref{Pe} above represents the transient saliency power.
\subsection{Available PMU measurements}
\lab{subsec22}
%
Typically, the following PMU measurements are available at the terminal of the SG modeled by \eqref{SG}:
\begin{equation}
	y=\col(\theta_t, V_t, \phi_t, I_t, P_t, Q_t),
\label{y}
\end{equation}
where $\theta_t(t)\in[0,2\pi)$ is the terminal voltage phase angle, $V_t(t)\in\rea_{>0}$ the terminal voltage magnitude, $\phi_t(t)\in[0,2\pi)$ the terminal current phase angle and $I_t(t)\in\rea_{>0}$ the terminal current magnitude as well as $P_t(t)\in\rea$ and $Q_t(t)\in\rea$ the active and reactive terminal power, respectively. 

Next, we establish five useful relations between the measurable PMU signals \eqref{y} and the states of the model \eqref{SG}.

\subsubsection{Expression for terminal voltages $V_q$ and $V_d$}
The terminal quadrature and direct axis voltages $V_q$ and $V_d$ can be expressed in the local machine $dq$-frame as \cite{sauer2017,MACetal}
\begin{equation}
\begin{bmatrix}
	V_q \\ V_d
\end{bmatrix}=\begin{bmatrix}
V_t \cos(x_1-\theta_t)\\
V_t \sin(x_1-\theta_t)
\end{bmatrix}=\begin{bmatrix}
y_2\cos(x_1-y_1)\\
y_2 \sin(x_1-y_1)
\end{bmatrix}.
\label{VqVd}
\end{equation} 
Then, by introducing the stator resistance $R\in\rea_{>0}$, the following relation between the terminal voltages $V_q$ and $V_d$ and the internal voltages $E_q'$ and $E_d'$, respectively, can be established, see \cite{sauer2017,MACetal}:
\begin{equation}
	\begin{bmatrix}
		V_q\\V_d
	\end{bmatrix} =\begin{bmatrix}
		E_q'\\E_d'
	\end{bmatrix} -\begin{bmatrix}
		R & X_d'\\-X_q' & R
	\end{bmatrix}\begin{bmatrix}
		I_q\\I_d
	\end{bmatrix},
\label{VdqEdq}
\end{equation}
or, equivalently,
\begin{equation}
	\begin{bmatrix}
	y_2\cos(x_1-y_1)\\
	y_2 \sin(x_1-y_1)
	\end{bmatrix} =\begin{bmatrix}
		x_3\\x_4
	\end{bmatrix} -\begin{bmatrix}
		R & X_d'\\-X_q' & R
	\end{bmatrix}\begin{bmatrix}
		I_q\\I_d
	\end{bmatrix}.
	\label{VdqEdq2}
\end{equation}

\subsubsection{Expression for $y_4^2=I_q^2+I_d^2$}
Rearranging \eqref{VdqEdq} for $I_q$ and $I_d$ yields
\begin{equation}
	\begin{split} 
	\begin{bmatrix}
		I_q\\I_d
	\end{bmatrix}& =
\begin{bmatrix}
	R & X_d'\\-X_q' & R
\end{bmatrix}^{-1}\left(
\begin{bmatrix}
	E_q'\\E_d'
\end{bmatrix} -
	\begin{bmatrix}
	V_q\\V_d
\end{bmatrix}
\right)\\ 
&=\frac{1}{R^2+X_d'X_q'}\begin{bmatrix}
	R & -X_d'\\X_q' & R
\end{bmatrix}
\begin{bmatrix}
x_3-y_2\cos(x_1-y_1)\\x_4-y_2 \sin(x_1-y_1)
\end{bmatrix}.
\end{split}
\label{IqId}
\end{equation}
Hence, we obtain
\begali{
\nonumber
y_4^2=&I_t^2=I_q^2+I_d^2\\
\nonumber
=&
\frac{1}{(R^2+X_d'X_q')^2}\big[ 
R (x_3-y_2\cos(x_1-y_1))\\
\nonumber
&-X_d'(x_4-y_2 \sin(x_1-y_1))^2\\
\nonumber
&+R(x_4-y_2 \sin(x_1-y_1))\\
&+X_q'(x_3-y_2\cos(x_1-y_1))^2\big].
\lab{y4squ0}
} 

\subsubsection{Expression for $y_5$}
The instantaneous terminal active power $P_t(t)\in\rea$ is given by
\begin{equation}
y_5=P_t=V_qI_q+V_dI_d.
\label{Pt}
\end{equation}
By using the relation \eqref{VdqEdq}, we can express $P_t$ in \eqref{Pt} as
\begin{equation*}
y_5=E_q' I_q+E_d'I_d+(X_q'-X_d')I_dI_q-R(I_q^2+I_d^2),
\end{equation*}
which provides the following relation between the internal, non-measurable electric air-gap power $P_e$ in \eqref{Pe} and the measurable terminal power $P_t$:
\begin{equation}
	\begin{split} 
		P_e=P_t+R(I_q^2+I_d^2)=y_5+Ry_4^2,
	\end{split} 
\label{PePt}
\end{equation}
where the last term describes the stator resistance loss.

\subsubsection{Expression for $y_6$}
The instantaneous terminal reactive power $Q_t(t)\in\rea$ is given by
\begin{equation}
	y_6=Q_t=V_qI_d-V_dI_q.
	\label{Qt}
\end{equation}

\subsubsection{Expressions for $I_q$ and $I_d$}
The formulas for $I_q$ and $I_d$ in \eqref{IqId} are rather complex. Alternatively, from \eqref{Pt} and \eqref{Qt}, we have that
\begin{equation*}
\begin{bmatrix}
P_t \\ Q_t
\end{bmatrix}=\begin{bmatrix} 
V_q& V_d\\
-V_d & V_q
\end{bmatrix} 
\begin{bmatrix}
I_q\\I_d
\end{bmatrix},
\end{equation*}
or, equivalently, with \eqref{VqVd}
\begin{equation}
	\begin{bmatrix}
		y_5 \\ y_6
	\end{bmatrix}=\begin{bmatrix} 
		y_2\cos(x_1-y_1)& y_2 \sin(x_1-y_1)\\
		-y_2 \sin(x_1-y_1) & y_2\cos(x_1-y_1)
	\end{bmatrix} 
	\begin{bmatrix}
		I_q\\I_d
	\end{bmatrix}.
\end{equation}
Rearranging for $I_q$ and $I_d$ yields
\begin{equation}
	\begin{bmatrix}
		I_q\\I_d
	\end{bmatrix}=\frac{1}{y_2}\begin{bmatrix}
	\cos(x_1-y_1) & -\sin(x_1-y_1)\\
	\sin(x_1-y_1)& \cos(x_1-y_1)
\end{bmatrix}	\begin{bmatrix}
y_5 \\ y_6
\end{bmatrix},
\label{IdIq2}
\end{equation}
where we have used
\begin{equation*}
\begin{split}
&\begin{bmatrix} 
		y_2\cos(x_1-y_1)& y_2 \sin(x_1-y_1)\\
		-y_2 \sin(x_1-y_1) & y_2\cos(x_1-y_1)
	\end{bmatrix}^{-1}
\\&=
\frac{1}{y_2}
\begin{bmatrix}
\cos(x_1-y_1) & -\sin(x_1-y_1)\\
\sin(x_1-y_1)& \cos(x_1-y_1)
\end{bmatrix},
\end{split}
\end{equation*}
which is well-defined since $y_2(t)\in\rea_{>0}$ (as it denotes the terminal voltage magnitude).
%
\subsection{Final model and problem formulation}
\lab{subsec23}

\begpro  \em
\lab{pro1}
Defining the constants
\begalis{
a_0&:=\frac{D\omega_0}{2H},\;a_1:=\frac{1}{T_{d0}'},\;a_2:=\frac{1}{T_{q0}'},\;b_0:=\frac{\omega_0}{2H},\\
 b_1&:= \frac{1}{T_{d0}'}(X_d-X_d'),\;b_2=\frac{1}{T_{q0}'}(X_q-X_q'),\; b_3:=\frac{\omega_0 R}{2H},
}
we finally can express \eqref{SG} using \eqref{PePt} and \eqref{IdIq2} as
\begsubequ
\lab{finmod}
\begali{
\lab{dotx1}
\dot x_1&=x_2,\\
\lab{dotx2}	
\dot x_2&=-a_0x_2+b_0(u_1-y_5)-b_3y_4^2,\\
\lab{dotx3}	
\dot x_3&=-a_1x_3-\frac{b_1}{y_2}(\sin(x_1-y_1)y_5+\cos(x_1-y_1)y_6)+a_1u_2,\\
\lab{dotx4}		
\dot x_4&=-a_2x_4+\frac{b_2}{y_2}(\cos(x_1-y_1)y_5-\sin(x_1-y_1)y_6).
}
\endsubequ
\endpro

\noindent {\bf State Observation Problem Formulations} Our aim to design a {\em state observer} for the system \eqref{finmod} with PMU measurements \eqref{y} of either one of the following forms.
\begenu[{\bf PF1}]
\item ({\em Complete} input measurement)
 \begalis{
	\dot \chi &= G_c(\chi,u,y)\\
	\hat x&=H_c(\chi_c,u,y),
}
with $\chi(t) \in \rea^{n_\chi}$, $G_c : \rea^{n_\chi} \times \rea^2 \times \rea^5 \to \rea^{n_\chi}$, $H_c : \rea^{n_\chi} \times \rea^2 \times \rea^5 \to \rea^{4}$
\item ({\em Partial} input measurement)
 \begalis{
	\dot \chi &= G_p(\chi,u_1,y)\\
	\hat x&=H_p(\chi,u_1,y),
}
\endenu
with $\chi(t) \in \rea^{n_\chi}$, $G_p : \rea^{n_\chi} \times \rea \times \rea^5 \to \rea^{n_\chi}$, $H_p : \rea^{n_\chi} \times \rea \times \rea^5 \to \rea^{4}$.

The objective in both scenarios is to ensure that  for all $x(0) \in \rea^4$, $\chi(0) \in \rea^{n_\chi}$ and all continuous $u(t)$ that generate a bounded state trajectory $x(t)$, we ensure that
\begalis{
&\lim_{t\to\infty} |\hat x(t)-x(t)|=0,
}
exponentially fast.

\begrem
\lab{rem1}
Notice that in the Problem Formulation {\bf PF1} we assume both inputs $u_1$ and $u_2$---that is, mechanical power $P_m$ and field voltage $E_f$---are available for measurement. On the other hand, in the Problem Formulation {\bf PF2} we assume only the mechanical power $P_m$ is measurable. This is a practically relevant scenario.
\endrem

\begrem
\lab{rem2}
It is interesting to compare the model above with the one used in \cite[Equation (1)]{BOBetaltac22}, obtained neglecting the rotor saliency and resistance, that is, setting $x_d'=x_q'$ and $R=0$. Namely 
\begali{
\nonumber
\dot x_1&=x_2,\\	
\nonumber
\dot x_2&=-a'_0x_2+b'_0(u_1-y_5),\\	
\nonumber
\dot x_3&=-a'_1x_3+b'_1y_2\cos(x_1-y_1)+c_1u_2,\\
\lab{simmod}
\dot x_4&=-a'_2x_4+b'_2y_2\sin(x_1-y_1),
}
where the physical definition of the new coefficients $(\cdot)'_i$ and $c_1$ is given in \cite{BOBetaltac22}. The simplification of the dynamic model when invoking the assumptions can hardly be overestimated. 
\endrem
%
\section{State Observer with Full Input Measurement}
\lab{sec3}
%
In this section we present the solution to Problem Formulation {\bf PF1}. First, in Subsection \ref{subsec31}, we enunciate  four technical lemmata that are instrumental for the solution of the problem. The first one provides a suitable parameterization for $x_1$ and constructs (up to an exponentially decaying term) $x_2$. In Lemma \ref{lem2} we give a linear parameterization of the transcendental functions appearing in the system dynamics \eqref{finmod}. In Lemma \ref{lem3} we give a parameterization of the subsystem $(x_3,x_4)$ that can be directly applied to derive a suitable regression equation for the parameter estimation, a task that is carried-out in Subsection \ref{subsec32}. Finally, in Subsection \ref{subsec33} we present the parameter and state observer derived using the LS+DREM parameter estimator.
%
\subsection{Preliminary lemmata}
\lab{subsec31}
%
\beglem
\lab{lem1}\em
Consider the subsystem \eqref{dotx1}, \eqref{dotx2}. Define the dynamic extension
\begalis{
\dot v_1&=v_2,\;v_1(0)=0\\
\dot v_2&=-a_0v_2+b_0(u_1-y_5)-b_3y_4^2,\;v_2(0)=0.
}
Then
\begsubequ
\lab{xv}
\begali{
\lab{x1v1}
x_1 &=v_1+\theta_0\\
\lab{x2v2}
x_2&=v_2 + \et
}
\endsubequ
where $\theta_0$ is an {\em unknown} constant parameter and $\et$ denotes a signal converging to zero exponentially fast
\endlem
\begproo
Define the error signals 
$$
\begmat{\tilde x_1 \\ \tilde x_2}:=\begmat{ x_1-v_1 \\ x_2-v_2},
$$
which clearly satisfy
$$
\begmat{\dot {\tilde x}_1 \\ \dot {\tilde x}_2}:=\begmat{ \tilde x_2 \\ -a_0 \tilde x_2}.
$$
Solving this differential equation we get
\begalis{
x_1(t) &=v_1(t)+x_1(0)+x_2(0)\int_0^t e^{-a_0 \tau}d\tau \\
x_2(t)&=v_2(t)+e^{-a_0 t}x_2(0).
}
The proof is completed setting $\et:=e^{-a_0 t}x_2(0)$, evaluating $x_1(t)$ at $t=0$ in the equation above, noting that $v_1(0)=0$, and defining $\theta_0:=x_1(0)$.\footnote{It will become clear in the sequel that $x_2$ does not appear in other calculations, consequently the presence of the exponentially decaying term in \eqref{x2v2} does not affect the claims regarding the asymptotic behavior of the signals.} 
\endproo

\beglem
\lab{lem2}\em
Consider the subsystem \eqref{dotx3}, \eqref{dotx4} and the identity \eqref{x1v1}. Then
\begequ
\lab{sincos}
\begmat{
\sin(x_1-y_1) \\ \cos(x_1-y_1)}=e^{-\calj(v_1-y_1)}\begmat{\theta_1 \\ \theta_2},
\endequ
where $\calj:=\begmat{0 & -1 \\ 1 & 0}$ and $\theta_1$, $\theta_2$ are {\em unknown} constant parameters.
\endlem
\begproo
Replacing \eqref{x1v1} in  \eqref{dotx3} and \eqref{dotx4} we get
\begali{
\nonumber
&\begmat{\sin(x_1-y_1) \\ \cos(x_1-y_1)}  =\begmat{\sin(v_1-y_1+\theta_0) \\ \cos(v_1-y_1+\theta_0)}\\
\nonumber
& =\begmat{\sin(v_1-y_1)\cos(\theta_0) +\cos(v_1-y_1) \sin(\theta_0)\\ \cos(v_1-y_1)\cos(\theta_0) -\sin(v_1-y_1) \sin(\theta_0)} \\
\nonumber
& = \begmat{\cos(v_1-y_1) & \sin(v_1-y_1) \\ -\sin(v_1-y_1) & \cos(v_1-y_1)} \begmat{ \sin(\theta_0)\\ \cos(\theta_0) }\\
\lab{sincos1}
& = e^{-\calj(v_1-y_1)}\begmat{\theta_1 \\ \theta_2},
}
where we defined 
\begali{
\lab{the1the2}
\begmat{\theta_1 \\ \theta_2}:=\begmat{\sin(\theta_0) \\ \cos(\theta_0)}.
}
\endproo

\beglem
\lab{lem3}\em
The following {\em linear parameterization} of the states $(x_3,x_4)$ holds
\begequ
\lab{x3x4}
\begin{bmatrix}x_3\\x_4\end{bmatrix} =W(v,y) \begmat{\theta_1 \\ \theta_2}, 
\endequ
where $v=\col(v_1,v_2)$ and $W(v,y) \in \rea^{2 \times 2}$ is {\em measurable}.
\endlem
\begproo
From \eqref{VdqEdq2}, we have that
\begin{equation}
\lab{x3x41}
\begin{bmatrix}
	x_3\\x_4
\end{bmatrix} =	\begin{bmatrix}
		y_2\cos(x_1-y_1)\\
		y_2 \sin(x_1-y_1)
	\end{bmatrix}+\begin{bmatrix}
	R & x_d'\\-x_q' & R
\end{bmatrix}\begin{bmatrix}
I_q\\I_d
\end{bmatrix}.
\end{equation}
Replacing $I_q$ and $I_d$ with \eqref{IdIq2} yields\\
\begin{equation}
	\begin{split} 
	\begin{bmatrix}
		x_3\\x_4
	\end{bmatrix} &=	\begin{bmatrix}
		y_2\cos(x_1-y_1)\\
		y_2 \sin(x_1-y_1)
	\end{bmatrix}\\
	&+\begin{bmatrix}
		R & x_d'\\-x_q' & R
	\end{bmatrix}\frac{1}{y_2}\begin{bmatrix}
		\cos(x_1-y_1) & -\sin(x_1-y_1)\\
		\sin(x_1-y_1)& \cos(x_1-y_1)
	\end{bmatrix}	\begin{bmatrix}
		y_5 \\ y_6
	\end{bmatrix}\\
&=y_2\begin{bmatrix}
0 & 1\\
1&0
\end{bmatrix}\begin{bmatrix}
\sin(x_1-y_1)\\
\cos(x_1-y_1)
\end{bmatrix}\\
& +\begin{bmatrix}
	R & x_d'\\-x_q' & R
\end{bmatrix}\frac{1}{y_2}\begin{bmatrix}
-y_6 & y_5\\
y_5 & y_6
\end{bmatrix}
\begin{bmatrix}
	\sin(x_1-y_1)\\
\cos(x_1-y_1)
\end{bmatrix}\\
&=\left(y_2\begin{bmatrix}
0 & 1\\
1&0
\end{bmatrix}
+\begin{bmatrix}
R & x_d'\\-x_q' & R
\end{bmatrix}\frac{1}{y_2}\begin{bmatrix}
-y_6 & y_5\\
y_5 & y_6
\end{bmatrix} 
\right)\times\\
& \quad e^{-\calj(v_1-y_1)}\begmat{\theta_1 \\ \theta_2},
\end{split} 
\end{equation}
where we have used \eqref{sincos1} in the last identity. The proof is completed with the definition  
\begin{equation}\begin{split}
&W(v,y):=\\&{1 \over y_2} \begmat{-R y_6+x_d'y_5 & R y_5+x_d'y_6+y_2^2 \\
R y_5+x_q'y_6 + y_2^2 & R y_6-x_q'y_5}e^{-\calj(v_1-y_1)}.
\end{split}\end{equation}
\endproo

\begrem
\lab{rem3}
A potential problem with Lemma \ref{lem1} is that it implements an {\em open-loop} integration of the system dynamics. It is well-known that this operation is sensitive to noise. 
\endrem
%
\subsection{State parameterization and corresponding regression equation}
\lab{subsec32}
%
From the calculations above we see that if we can {\em estimate $\theta_0$} the state observation problem is solved. Indeed, if $\theta_0$ is known we can compute $x_1$ from \eqref{x1v1}. The state $x_2$ is given by \eqref{x1v1}. Moreover, knowing $\theta_0$, we can compute $(\theta_1,\theta_2)$ using \eqref{the1the2}. Finally, replacing the latter in \eqref{x3x4} we obtain $(x_3,x_4)$. Unfortunately, with the available PMU measurements \eqref{y}, it does not seem possible to construct a {\em regression equation} for $\theta_0$ that we could use in a parameter estimator. 

In this subsection we present a solution to this problem, utilizing \eqref{x3x4} and \eqref{y4squ} to construct a regression equation for $(\theta_1,\theta_2)$, from which we can compute $\theta_0$ via
\begequ
\lab{the0}
\theta_0 = \arctan\Big\{{\theta_1 \over \theta_2}\Big\},
\endequ
which follows directly from the definition \eqref{the1the2}. The regression equation is {\em nonlinearly parameterized} but, as indicated in the introduction we propose to use the new LS+DREM estimator of \cite{ORTROMARA} that can handle NLPRE.

The final design, that is, the definition of the LS+DREM parameter estimator, the state estimator, and the assessment of its convergence properties, is carried-out in the next subsection.

\begpro \em
\lab{pro2}
Consider the linear parameterization of the states $(x_3,x_4)$ defined in \eqref{x3x4} and the expression for $y_4$ given in  \eqref{y4squ}. The system satisfies the following NLPRE
\begequ
\lab{nlpre}
\caly = \psi_1^\top \begmat{\theta_1 \\ \theta_2}+\psi_2^\top \calg(\theta_1,\theta_2),
\endequ
where $\caly(t) \in \rea$, $\psi_1(t) \in \rea^{2}$ and $\psi_2(t) \in \rea^{4}$ are {\em measurable} signals and $\calg:\rea^2 \to \rea^{3}$ is the mapping
\begequ
\lab{calg}
\calg(\theta_1,\theta_2)=\col(\theta_1^2,\theta_2^2, \theta_1\theta_2).
\endequ
For the particular case of $R=0$ the NLPRE simplifies to
\begequ
\lab{nlpre1}
\caly = \psi_2^\top \calg(\theta_1,\theta_2),
\endequ
\endpro

\begproo
First, recall the definition of $y_4^2$ given in \eqref{y4squ} that, for ease of reference, we repeat here as:
\begali{
\nonumber
y_4^2=&\frac{1}{(R^2+X_d'X_q')^2} \big[ 
R (x_3-y_2\cos(x_1-y_1))\\
\nonumber
&-X_d'(x_4-y_2 \sin(x_1-y_1)))^2\\
\nonumber
&+R(x_4-y_2 \sin(x_1-y_1))\\
&+X_q'(x_3-y_2\cos(x_1-y_1)))^2\big].
\lab{y4squ}
}

The gist of the proof is to replace in the equation above the expressions of $(x_1,x_3,x_4)$ given in \eqref{sincos} of Lemma \ref{lem2} and \eqref{x3x4} and to group terms to obtain \eqref{nlpre}.  We recall here these equations:
\begsubequ
\lab{sincosx3x4}
\begali{
\lab{sin}
\sin(x_1-y_1)&= \cos(v_1-y_1)\theta_1+\sin(v_1-y_1)\theta_2\\
\lab{cos}
\cos(x_1-y_1)&= \cos(v_1-y_1)\theta_2-\sin(v_1-y_1)\theta_1\\
\begmat{x_3 \\ x_4} & =W(v,y) \begmat{\theta_1 \\ \theta_2},
\lab{eqs}
}
\endsubequ
which are  {\em linearly dependent} on $(\theta_1,\theta_2)$. It is clear then  that replacing \eqref{sincosx3x4} in \eqref{y4squ} yields an expression that contains terms {\em linearly} depending on $(\theta_1,\theta_2)$, as well as, {\em quadratic} terms on  $(\theta_1,\theta_2)$. Moreover, the linearly dependent terms {\em disappear} if $R=0$ and only the quadratic terms are retained. 

In the interest of brevity we present below the expressions {\em for the case $R=0$}. After some lengthy, but straightforward, calculations we identify the terms of the NLPRE \eqref{nlpre} as 
\begin{equation*}
	\begin{split}
		c_v &:= \cos(v_1-y_1), \quad s_v := \sin(v_1-y_1) ,\\
		a_1 &:= X_d'y_5c_v - \left(X_d'y_6+y_2^2\right)s_v,\\
		a_2 &:= X_d'y_5s_v +\left(X_d'y_6+y_2^2\right)c_v,\\
		a_3 &:= \left(X_q'y_6+ y_2^2 \right)c_v +X_q'y_5s_v,\\
		a_4 &:= \left(X_q'y_6+ y_2^2 \right)s_v - X_q'y_5c_v,\\
			\psi_{2,1} &:= 	\frac{1}{(X_d')^2}\left( a_1^2/y_2^2  + s_v^2y_2^2  + 2a_1s_v\right) \\& \ \ \ \  +\frac{1}{(X_q')^2}\left( a_3^2/y_2^2 + c_v^2y_2^2 - 2a_3c_v\right),\\
				\psi_{2,2} &:= 	\frac{1}{(X_d')^2}\left(a_2^2/y_2^2 + c_v^2y_2^2 - 2a_2c_v\right) 
			\\& \ \ \ \ 	+\frac{1}{(X_q')^2}\left(  a_4^2/y_2^2   + s_v^2y_2^2  - 2a_4s_v\right), \\
					\psi_{2,3} &:= 	2 \bigg(  \frac{1}{(X_d')^2}\left( -a_1c_v + a_2s_v  + a_1a_2/y_2^2  -c_vs_vy_2^2 \right) 
				\\& \ \ \ \ 	+\frac{1}{(X_q')^2}\left(  -a_4c_v  -a_3s_v + a_3a_4/y_2^2 + c_vs_vy_2^2\right) \bigg),\\
		\mathcal{Y} &:= 	y_4^2 ,\quad
		\psi_1 := \begin{bmatrix}
			0 &0
		\end{bmatrix}^\top, \quad 		\psi_2 :=\begin{bmatrix}
			\psi_{2,1} &
			\psi_{2,2}&
		\psi_{2,3}
		\end{bmatrix}^\top.
	\end{split}
\end{equation*}
It is clear from  \eqref{y4squ} that the presence of terms appearing with $R \neq 0$ introduces terms {\em linearly dependent} on $(\theta_1,\theta_2)$ that will be captured by the vector $\psi_1$ of the NLPRE \eqref{nlpre}, with the same $\psi_2$ and $\calg(\theta_1,\theta_2)$. 
\endproo
%
\subsection{Parameter and state estimator}
\lab{subsec33}
%
We present in this subsection the parameter estimator that is needed to complete our observer. The proof of the proposition may be found in \cite{ORTROMARA}. To streamline the proposition given below we need the following.

\begass 
\lab{ass1} \em
 Consider the NLPRE \eqref{nlpre}. The regressor vector $\psi$ defined as
\begequ
\lab{psi}
\psi:=\begmat{\psi_1 \\ \psi_2},
\endequ
is IE \cite{KRERIE}, \cite[Definition 3.3]{TAObook}. That is, there exists $t_c>0$ and $\delta>0$ such that
$$
\int_0^{t_c} \psi(s)\psi^\top(s)ds > \delta I_{3}.
$$
\endass

We recall that in \cite[Lemma 3]{WANORTBOB} it is shown that for the single output linear regression equation case this is {\em equivalent} to the existence of a time sequence $\{t_j\}_{j=1}^3$ such that 
$$
\rank\Bigg\{\begmat{\psi^\top(t_1)\\ \psi^\top(t_2)\\ \psi^\top(t_{3})}\Bigg\}=3,
$$
which is the definition of (off- or on-line) {\em identifiability} \cite{GOOSINbook}.

\begpro
\lab{pro4}\em
Consider the system \eqref{finmod} and the associated regression equation \eqref{nlpre} compactly rewritten as
$$
\caly=\psi^\top \calg_0(\theta_1,\theta_2),
$$
with 
$$
\calg_0(\theta_1,\theta_2):=\begmat{\theta_1 \\ \theta_2 \\ \calg(\theta_1,\theta_2)},
$$
and $\psi$, given in \eqref{psi},  {\em bounded} and verifying {\bf Assumption} \ref{ass1}. Assume we know a {\em positive constant} $\theta_2^a$ such that the initial condition of the rotor angle verifies
\begequ
\lab{boucosx1}
-\theta_2^{a} \geq \cos(x_1(0)) \geq \theta_2^{a}.
\endequ
Define the LS+DREM estimator with forgetting factor \cite{ORTROMARA}.
\begali{
\nonumber
		\dot{\hat \calg}_0 & =\gamma_\calg F \psi (\caly-\psi^\top \hat\calg_0),\; \hat\calg_0(0) \in \rea^{5}\\
\nonumber
\dot {F}& =  -\gamma_\calg F \psi \psi^\top F +\chi F,\; F(0)={1 \over f_0} I_{5} \\
\nonumber
\dot{\hat \theta}_1 & = \gamma_1  \Delta [{\bf Y}_1 -\Delta \hat\theta_1],\; \hat\theta_1(0) \in \rea \\
\nonumber
\dot{\hat \theta}_2 & = \gamma_2  \Delta [{\bf Y}_2 -\Delta \hat\theta_2 ],\;-\theta_2^{a} \geq \hat \theta_2(0) \geq \theta_2^{a} \\
\nonumber
\dot z &=\; -\chi z, \; z(0)=1, \\
\chi &= \chi_0 \left( 1-{{\| F\|}\over{k}} \right)
\lab{lsd}
}
with tuning gains the scalars $\gamma_\calg>0, \gamma_i>0$, $f_0>0$, $\chi_0> 0$ and $k\geq \frac{1}{f_0}$, and we defined the signals
	\begalis{
		\Delta & :=\det\{I_{5}-z f_0F\}\\
		{\bf Y} & := \adj\{I_{5}- zf_0F\} (\hat\calg_0 -  zf_0F \calg_{0}),
	}
where $ \adj\{\cdot\}$ denotes the adjugate matrix.
\begenu
\item [{\bf (i)}] For all initial conditions the estimated parameters verify
$$
\liminf\Big|\begmat{\hat \theta_1(t) \\ \hat \theta_2(t)}- \begmat{\theta_1 \\ \theta_2}\Big|=0,
$$
exponentially fast.
\item [{\bf (ii)}] The estimate $\hat \theta_2(t)$ is bounded away from zero for all $t \geq 0$. Moreover it satisfies the bounds
$$
-\theta_2^{a} \geq \hat \theta_2(t) \geq \theta_2^{a}.
$$
\item [{\bf (iii)}] The state estimation 
\begalis{
\hat x_1 & = v_1+ \arctan\Big\{{\hat \theta_1 \over \hat \theta_2}\Big\} \\
\hat x_2 & = v_2\\
\begmat{ \hat x_3\\ \hat x_4 }& =W(v,y) \begmat{ \hat \theta_1 \\ \hat \theta_2}, 
}
verifies 
\begequ
\lim_{t\to\infty} |\hat x(t)- x(t)|=0,
\lab{expcon}
\endequ
exponentially fast.

\item [{\bf (iv)}] All the signals are {\em bounded}.
\endenu 
\endpro

\begrem
\lab{rem4}
The proof of the claim {\bf (ii)} follows for the well-known fact that DREM estimators ensure {\em monotonicity} of the parameter estimation errors \cite[Corollary 1]{ORTROMARA}. 
\endrem

\begrem
\lab{rem5}
In this paper we proposed to use the LS+DREM estimator because of its ability to deal with NLPRE and the weak excitation requirements imposed on the regressor to ensure global, exponential convergence. However, there are other estimators---for instance, the GPEBO+DREM estimator of \cite{WANetalijc} that enjoy the same features. Moreover, in a scenario where "good excitation" is available, it is possible to overparameterize the NLPRE to generate a LRE, and still achieve parameter convergence with a standard gradient or LS estimator. 
\endrem
%
\section{State Observer with Partial Input Measurement}
\lab{sec4}
%
To streamline the presentation of the main result we define the following signals:
\begsubequ
\lab{w}
\begali{
\lab{w1}
w_1&:={1 \over y_2} e^{\calj(v_1-y_1)}\begmat{-y_6 \\ y_5}\\
\lab{w2}
w_2&:={1 \over y_2} e^{\calj(v_1-y_1)}\begmat{y_5 \\ y_6}\\
\lab{w3}
w_3&:= e^{\calj(v_1-y_1)}\begmat{y_2 \\ 0}- x'_q w_1+R w_2,
}
\endsubequ
where, we recall, that $v_1$ is defined in Lemma \ref{lem1}.
 
\begpro
\lab{pro5}\em
Consider the SG model \eqref{finmod} with {\em partial input measurement} $u_1=P_m$. Define the dynamic extension
\begequ
\lab{xi}
\dot \xi=-a_2 \xi + b_2 w_1,
\endequ
with $w_1(t) \in \rea^2$ given in \eqref{w1}. Define the estimates
\begsubequ
\lab{hatthe}
\begali{
\lab{hatthe0}
\hat \theta_0 &:=\arctan\Bigg\{ {\xi_2 - w_{3,2} \over w_{3,1}-\xi_1} \Bigg\}\\
\lab{hatthe12}
\begmat{\hat \theta_1 \\ \hat \theta_2}&:=\begmat{\sin(\hat \theta_0) \\ \cos(\hat \theta_0)},
}
\endsubequ
where $w_3(t)=\col(w_{3,1}(t),w_{3,2}(t)) \in \rea^2$ is defined in \eqref{w3}.

\begenu
\item [{\bf (i)}] For all initial conditions the state estimation 
\begalis{
\hat x_1 & = v_1+\hat \theta_0 \\
\hat x_2 & = v_2 \\
\begmat{ \hat x_3\\ \hat x_4 }& =W(v,y) \begmat{ \hat \theta_1 \\ \hat \theta_2}, 
}
verifies \eqref{expcon}.
\item [{\bf (ii)}] All the signals are {\em bounded}.
\endenu 
\endpro

\begproo
From the second equation in \eqref{x3x41} we have
\begequ
\lab{x42}
x_4=y_2 \sin(x_1-y_1)+\begmat{-x_q' & R}\begin{bmatrix}I_q\\I_d\end{bmatrix}.
\endequ
On the other hand, from \eqref{IdIq2}
\begalis{
	\begmat{I_q\\I_d}&=\frac{1}{y_2}e^{\calj(x_1-y_1)}	\begmat{y_5 \\ y_6}\\
	&=\frac{1}{y_2}\begmat{\begmat{\theta_1 & \theta_2}e^{\calj(v_1-y_1)}\calj \\ \begmat{\theta_1 & \theta_2}e^{\calj(v_1-y_1)}}\begmat{y_5 \\ y_6}\\
	&=\frac{1}{y_2}\begmat{\begmat{-y_6 & y_5} e^{-\calj(v_1-y_1)}\\ \begmat{y_5 & y_6}e^{-\calj(v_1-y_1)}}\begmat{\theta_1 \\ \theta_2}\\
	&=\begmat{w_1^\top\\ w_2^\top}\begmat{\theta_1 \\ \theta_2},
}
where we used \eqref{sincos} to get the second identity, and the definitions \eqref{w} in the last one. Replacing the equation above in \eqref{x42} we get
\begali{
\nonumber
x_4 & =y_2 \sin(x_1-y_1)+\begmat{-x_q' & R}\begmat{w_1^\top\\ w_2^\top}\begmat{\theta_1 \\ \theta_2}\\
\nonumber
& =y_2 \begmat{1 & 0}e^{-\calj(v_1-y_1)}\begmat{\theta_1 \\ \theta_2}+(-x_q'w_1+ Rw_2)^\top\begmat{\theta_1 \\ \theta_2}\\
\lab{firx4}	
&= w_3^\top\begmat{\theta_1 \\ \theta_2}.
}

Now, replacing \eqref{IdIq2} in \eqref{dotx4} we get
\begalis{
\dot x_4 & =-a_2 x_4+b_2I_q\\
  & =-a_2 x_4+b_2 w_1^\top \begmat{\theta_1 \\ \theta_2},
 }
that we can write as
\begequ
\lab{secx4}
x_4=\xi^\top \begmat{\theta_1 \\ \theta_2},
\endequ
where $\xi$ is defined in \eqref{xi}. Equating \eqref{firx4} and \eqref{secx4}, and rearranging terms we get \eqref{hatthe0}. The definition \eqref{hatthe12} is motivated by  \eqref{the1the2}. Finally, the definition of the estimated states follows from \eqref{x1v1} and \eqref{eqs}, respectively. 
\endproo
%
\section{Simulation Results}
\lab{sec5}

\begin{figure}
	\centering
	\includegraphics[width=1\linewidth]{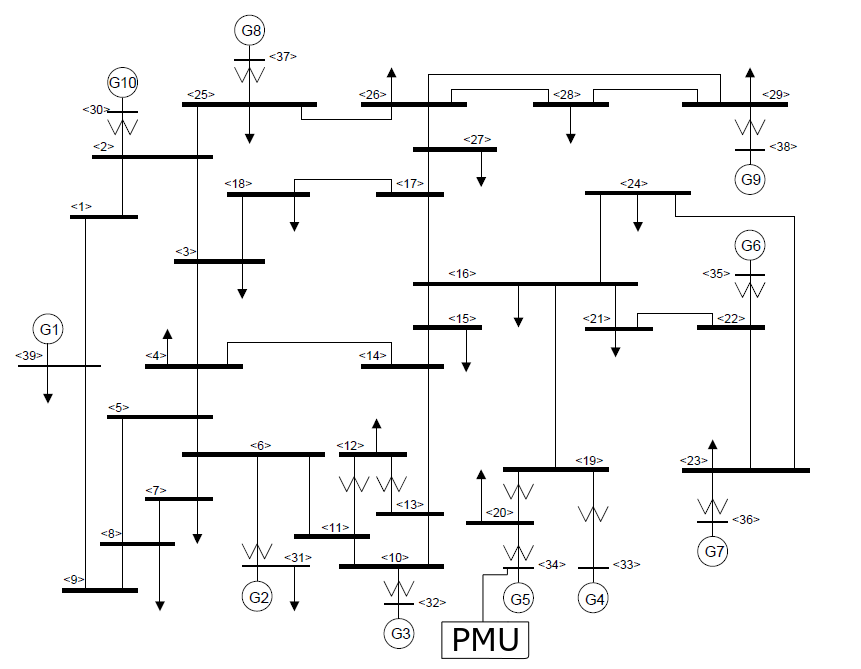}
	\caption{New England IEEE 39 bus system (figure taken from \cite{hiskens_ieee_2013}).}
	\label{fig:IEEE39Bus}
\end{figure}

In this section, we present simulation results to demonstrate the effectiveness of the proposed state observation scheme. For this, we utilize the well-known IEEE New England 39-bus system seen in Figure \ref{fig:IEEE39Bus} and described in detail in \cite{CANetal}.
As per \cite{hiskens_ieee_2013}, all synchronous generators are equipped with automatic voltage regulators (AVRs) and power system stabilizers (PSSs) and are modeled using the fourth-order model \eqref{SG}. Consequently, the following 9-dimensional model represents each generator:
\begin{equation*}
	\label{eq:full_sg}
	\begin{split}
		\dot x_1&=x_2,\\
		\dot x_2&=\frac{\omega_0}{2H}\left(u_1-P_e-D x_2 \right),\\
		\dot x_3&=\frac{1}{T_{d0}'}\left(u_2-x_3-(X_d-X_d')I_d \right),\\
		\dot x_4&=\frac{1}{T_{q0}'}\left(-x_4+(X_q-X_q')I_q \right),\\
		\dot{q} &= \frac{1}{T_{B}} \left(\left(1-\frac{T_{C}}{T_{B}}\right)(V_{\text{ref}} - V_{f}+ V_{\text{pss}})-q\right),\\
		\dot{E}_{f} &= \frac{1}{T_{A}} \left(K_{A}\left(q+\frac{T_{C}}{T_{B}}\right)(V_{\text{ref}} - V_{f}+ V_{\text{pss}})-E_{f}\right),\\
		\dot{p}_{1} &= -c_{1}p_{1} + p_{2} + (c_{4} -c_{1}c_{3})x_{2},\\
		\dot{p}_{2} &= -c_{2}p_{1} + p_{3} + (c_{5} -c_{2}c_{3})x_{2},\\
		\dot{p}_{3} &= -p_{1} -c_{1}c_{3}x_{2},\\
		V_{\text{pss}} &=p_{1} + c_{3,1} x_{2}, 
	\end{split}
\end{equation*}
where the intermediate variables needed for the AVR and PSS are $V_f$, $q$, and $p_j$, $j = 1, 2, 3$,
the differential equations for the PSS and AVR are according to \cite[Figs. 2, 3]{hiskens_ieee_2013} and where we defined the following constants
\begin{equation*}
	\begin{split}
		c_{1} & = \frac{T_{4}T_{w}+T_{4}T_{2} + T_{2}T_{w} }{T_{w} T_{4} T_{2}} ,\quad
		c_{2}  = \frac{T_{w} +T_{4} +T_{2}}{T_{w} T_{4} T_{2}},\\
		c_{3} & = \frac{K_{p} T_{1} T_{3}} {T_{2} T_{4}},\quad
		c_{4}  = \frac{K_{p} (T_{1}+ T_{3})} {T_{2} T_{4}},\quad
		c_{5}  =\frac{K_{p}}{T_{2} T_{4}}.
	\end{split}
\end{equation*}

We assume that a PMU is installed at the terminal bus of generator 5 to monitor the system. However, using the suggested approach and assuming additional PMUs are installed at the terminal buses of any generators, monitoring any or all other generators in the system is feasible. In order to replicate realistic PMUs, the measurements' sample frequency is set to 60 Hz, which is within the common PMU sampling rate range of 10-120 Hz \cite{zhou2014dynamic} and corresponds to the nominal system frequency.

To validate the observation technique presented in Proposition~\ref{pro5}, we simulate small load fluctuations in the system. The resulting frequency changes are consistent with those that occur during typical transmission grid operation \cite{weissbach2009verbesserung}. 

\begin{figure}
	\centering
	\includegraphics[width=0.5\textwidth]{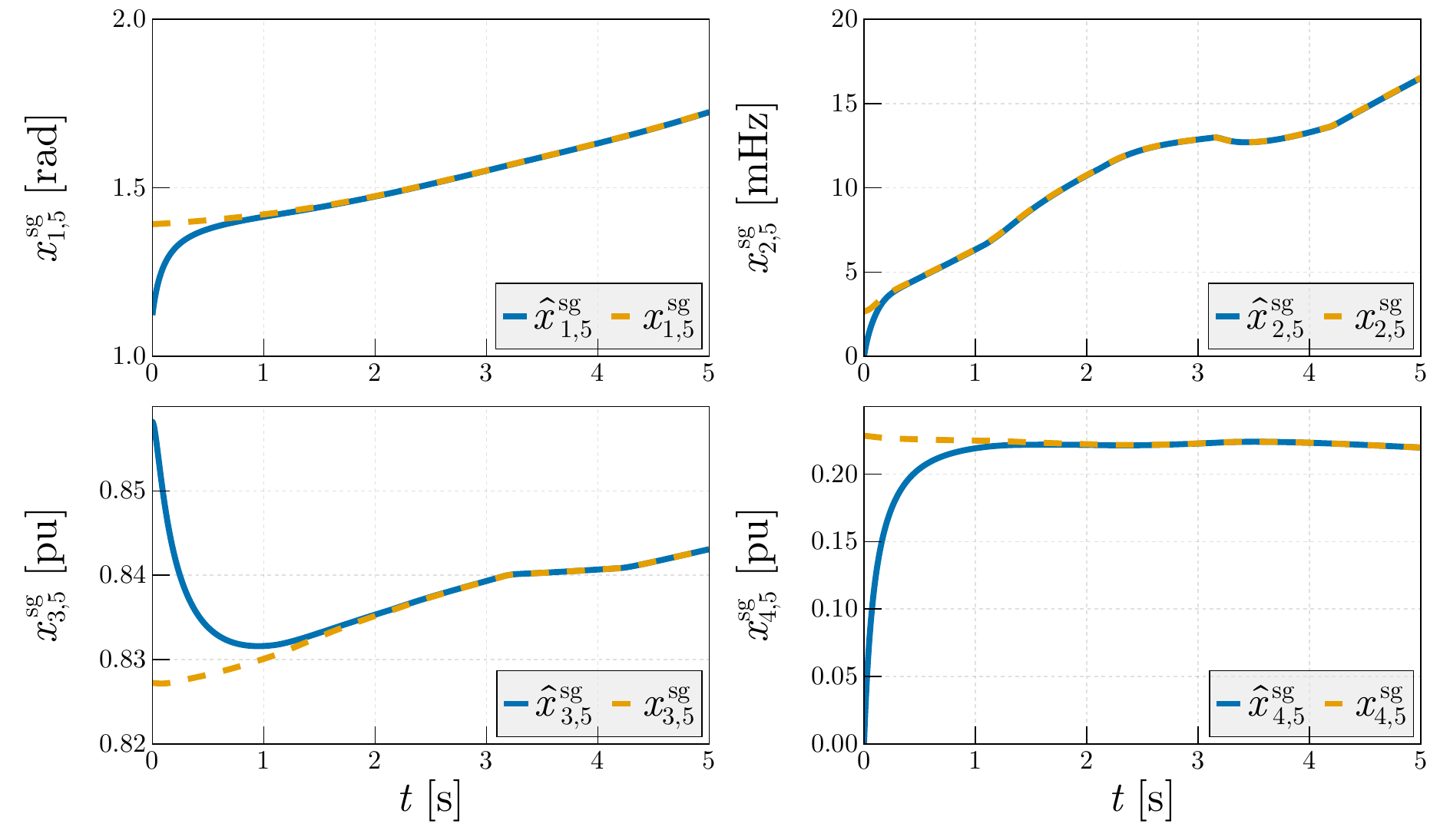}
	\caption{Transients of the state observer with partial input measurement of Proposition \ref{pro5} for generator 5 in the presence of load variations.}
	\label{fig:sim_pro4}
\end{figure}

In Figure~\ref{fig:sim_pro4}, the transients of the state observer with partial input measurement of Proposition \ref{pro5} for generator 5 are shown. It can be seen that the observation scheme is able to accurately reconstruct the states of the generator after a very short time frame of only around one second. Hence, it is well suited to monitor the dynamic states of synchronous generators in real-time. 
%
\section{Concluding Remarks and Future Research}
\lab{sec6}

We have extended the results of our previous work \cite{BOBetaltac22} on observer design for the state estimation, from PMU measurements, of multimachine power systems described by the widely popular fourth order model \eqref{SG}, in the following directions.
\begenu[{\bf E1}]
\item Relaxed the assumptions that the transient saliency is {\em neglected} and that the generators {\em stator resistance is zero}.
\item Design an observer where only mechanical power is available for measurement.
\endenu
 
Preliminary simulation results which small load fluctuations show the excellent performance of the second observer. In the final version of the paper we will present simulations of both observers in more realistic scenarios. 

\section*{Acknowledgment}
This paper is partially supported by the National Natural Science Foundation of China (61473183, U1509211), by the Ministry of Science and Higher Education of Russian Federation, passport of goszadanie no. 2019-0898.

N. Lorenz-Meyer and J. Schiffer acknowledge partial funding from the German Federal Government, the Federal Ministry of Education and Research, and the State of Brandenburg within the framework of the joint project EIZ: Energy Innovation Center (project numbers 85056897 and 03SF0693A).
		
\bibliographystyle{plain}
\bibliography{bib_JS}

\end{document}